\documentclass[twoside]{article}
\usepackage{qic,epsfig}
\usepackage{amsmath}
\usepackage{array}
\usepackage{graphicx}

\begin{document}
\setlength{\textheight}{8.0truein}    

\runninghead{Title  $\ldots$}
            {Author(s) $\ldots$}

\normalsize\textlineskip
\thispagestyle{empty}
\setcounter{page}{1}

\copyrightheading{0}{0}{2003}{000--000}

\vspace*{0.88truein}

\alphfootnote

\fpage{1}

\centerline{\bf
IMPROVED RECONCILIATION WITH POLAR CODES}
\vspace*{0.035truein}
\centerline{\bf IN QUANTUM KEY DISTRIBUTION}
\vspace*{0.37truein}
\centerline{\footnotesize
SUNGHOON LEE\footnote{sunghoon89@korea.ac.kr}}
\vspace*{0.015truein}
\centerline{\footnotesize\it School of Electrical Engineering, Korea University}
\baselineskip=10pt
\centerline{\footnotesize\it 145, Anam-ro, Seongbuk-gu, Seoul, 02841,
Republic of Korea}
\vspace*{10pt}
\centerline{\footnotesize 
JOOYOUN PARK, JUN HEO\footnote{junheo@korea.ac.kr}}
\vspace*{0.015truein}
\centerline{\footnotesize\it School of Electrical Engineering, Korea University}
\baselineskip=10pt
\centerline{\footnotesize\it 145, Anam-ro, Seongbuk-gu, Seoul, 02841,
Republic of Korea}
\vspace*{0.225truein}
\publisher{(received date)}{(revised date)}

\vspace*{0.21truein}

\abstracts{
Quantum key distribution (QKD) is a cryptographic system that generates an information-theoretically secure key shared by two legitimate parties. QKD consists of two parts: quantum and classical. The latter is referred to as classical post-processing (CPP). Information reconciliation is a part of CPP in which parties are given correlated variables and attempt to eliminate the discrepancies between them while disclosing a minimum amount of information. The elegant reconciliation protocol known as \emph{Cascade} was developed specifically for QKD in 1992 and has become the de-facto standard for all QKD implementations. However, the protocol is highly interactive. Thus, other protocols based on linear block codes such as Hamming codes, low-density parity-check (LDPC) codes, and polar codes have been researched. In particular, reconciliation using LDPC codes has been mainly studied because of its outstanding performance. Nevertheless, with small block size, the bit error rate performance of polar codes under successive-cancellation list (SCL) decoding with a cyclic redundancy check (CRC) is comparable to state-of-the-art turbo and LDPC codes. In this study, we demonstrate the use of polar codes to improve the performance of information reconciliation in a QKD system with small block size. The best decoder for polar codes, a CRC-aided SCL decoder, requires CRC-precoded messages. However, messages that are sifted keys in QKD are obtained arbitrarily as a result of  a characteristic of the QKD protocol and cannot be CRC-precoded. We propose a method that allows arbitrarily obtained sifted keys to be CRC precoded by introducing a virtual string. Thus the best decoder can be used for reconciliation using polar codes and improves the efficiency of the protocol.
}{}{}

\vspace*{10pt}

\vspace*{3pt}
\communicate{to be filled by the Editorial}

\vspace*{1pt}\textlineskip    

\section{Introduction}    
\label{Sec:Intro}
Quantum key distribution (QKD) is a secure means for two legitimate parties, known as Alice and Bob, to share secret keys between them. QKD consists of two parts: quantum and classical \cite{QuantumCryptography}. The quantum part involves the transmission of qubits through the quantum channel as well as their manipulation and detection. The classical part involves sifting, information reconciliation, and privacy amplification. Combining reconciliation and privacy amplification results in classical post-processing (CPP), which is performed using a public noiseless channel.

 A representative QKD protocol is the BB84 protocol proposed by Bennett and Brassard in 1984 \cite{BB84}. Alice sends qubits encoded in two orthogonal bases---diagonal and rectilinear---and Bob detects the qubits having arbitrary bases. Bob communicates those qubits he has detected and the measurements he has performed. Alice and Bob retain only the qubits with the right measurements and discard the rest. Therefore, the two possess a secret key that is half the length of the transmitted qubits.

In any realistic implementation of a QKD protocol, the secret keys of Alice and Bob can have discrepancies because of errors in the detector or the presence of an eavesdropper known as Eve. Therefore, every QKD protocol must include a CPP step in order for the parties to extract identical secret keys. Information reconciliation is the process of finding and correcting errors through public discussion \cite{BBBSS}. Reconciliation is performed using a public channel; therefore Eve can access information from it. The partial information that Alice can access, called leaked information, is the main performance indicator of a reconciliation protocol.

In a reconciliation protocol, information about a secret key is inevitably exposed to Eve. Thus, the two parties must perform an additional process called privacy amplification \cite{PA}. This is a method for reducing and effectively eliminating Eve's partial information about the secret key. The method can be performed using universal hash functions, which are implemented by the Toeplitz matrix, a construction for families of universal hash functions. The Toeplitz matrix can implement these functions efficiently with complexity $O(n\log{n})$ for input length $n$ \cite{Hayashi}.
The input of privacy amplification consists of many blocks that are error-free after reconciliation and verification.

The first reconciliation scheme was proposed in \cite{BBBSS}. It was a simple scheme containing a binary search algorithm. Afterwards, the very elegant and compact algorithm known as \emph{Cascade} was proposed in \cite{Cascade}. However, some problems such as high interactivity exist with \emph{Cascade}, several protocols have been developed that uses linear block codes. \emph{Winnow}, which uses Hamming codes, was introduced in 2003 \cite{Winnow}. A reconciliation protocol using low-density parity-check (LDPC) codes was proposed in \cite{Elkouss1}. 
The use of polar codes in reconciliation was proposed in \cite{Jouguet}. However, this reconciliation protocol employed a successive cancellation (SC) decoder proposed by Arikan and the size of a reconciliation block was too large. The performance of polar codes under an SC decoder was far behind that of state-of-the-art codes, and the size of a reconciliation block should be smaller. A reconciliation block is a unit which one codeword is applied to. Therefore, the code and the reconciliation block are the same. Unless specifically stated, the term ``block'' means a reconciliation block. 

In this study, we propose a reconciliation protocol that outperforms reconciliation protocols using LDPC codes with small block size. In previous experiments  \cite{experiment1,experiment2,experiment22,experiment3} on conducting a reconciliation, the size of a block did not exceed $6.6\times10^3$. Therefore, we focused on a reconciliation protocol with block sizes 2048 and 4096. The main idea of this study is based on a result that the performance of polar codes with length 2048 was better than that of state-of-the-art turbo and LDPC codes of similar length \cite{ListDecoding}. This result was achieved by using a cyclic redundancy check (CRC)-aided successive-cancellation list (SCL) decoder for polar codes. However, sifted keys of two parties are obtained arbitrarily because of a characteristic of the QKD protocol. Arbitrary sifted keys cannot be CRC-precoded. We propose a protocol that allows arbitrarily obtained sifted keys to be CRC-precoded by introducing a virtual string. Thus, the CRC-aided SCL decoder can be used for reconciliation using polar codes and increases the efficiency of the protocol.

The remainder of this paper is organized as follows. Section~\ref{Sec:Information Reconciliation} describes the preliminaries of this study, introduces previous research, and discusses information reconciliation in practical implementations. Section~\ref{Sec:Imroved Reconciliation} introduces the proposed protocol and discusses the application of polar codes under a CRC-aided SCL decoder to QKD. Section~\ref{Sec:Results} presents the practical implementation results and considers the complexity of protocols. Section \ref{Sec:Conclusions} provides a conclusion.

\section{Information Reconciliation}
\label{Sec:Information Reconciliation}

\subsection{Preliminaries}
\label{Subsec:Preliminaries}
Information reconciliation is a generic term to describe any method that can be used by two legitimate parties (Alice and Bob) to extract common information, provided that they have access to two correlated sequences, $X$ and $Y$ \cite{VanAssche}. In other words, $X$ and $Y$ are considered as the input and output of a modeled quantum channel, respectively.
It is commonly assumed that Alice and Bob hold sifted keys \textbf{x} and \textbf{y}, two $n$-length strings that are outcomes of $X$ and $Y$, which are the measurement results after the quantum part of QKD protocols is completed. During reconciliation, Alice and Bob communicate a set of messages $M$ to reconcile their keys. At the end of the process, they agree on a common string in the presence of an adversary while revealing some information. Therefore, $X$ and $Y$ can be considered correlated random variables, and every symbol in $Y$ is given by transition probability $p_{W}(\textbf{y}|\textbf{x})$ or equivalently is given as if every symbol was generated by memoryless channel $W$. In most QKD protocols, errors are usually uncorrelated and symmetric. For this reason, the memoryless channel $W$ can be seen as a binary symmetric channel (BSC). The crossover probability $p$ of the BSC is generally supposed as given.

The problem is how to encode $X$ and $Y$ into a message $M$. This encoding problem is known as Slepian-Wolf coding, which is a method of coding two lossless compressed correlated sources theoretically \cite{SlepianWolf}.
Because $(X,Y) \sim P_{X,Y}$ is an arbitrary pair of two correlated sequences over $\mathcal{X}\times\mathcal{X} $ that Alice and Bob have access to $\mathcal{X}=\lbrace 0,1 \rbrace$, we regard $(X,Y)$ as a memoryless source which are inputs and outputs of a quantum channel modeled by a BSC, respectively. Here, $X$ is the part to be compressed and $Y$ is the ``side information" about $X$.

Let $(X^n,Y^n)$ be an output measurement of length $n$ by the source.
The Slepian-Wolf bound states that $(X^n,Y^n)$ can be compressed into more than $nH(X|Y)$ bits without information loss. Therefore, given the codeword and side information $Y^n$, a decoder can recover $X^n$ without loss as $n$ approaches infinity.
When two correlated sequences $X$ and $Y$ are given, and $Y$ is only available at the decoder, the Slepian-Wolf theorem gives a theoretical bound for the lossless coding rate $R_X \geq H(X|Y)$. This is the
minimum information required to reconcile $X^n=\textbf{x}$ and $Y^n=\textbf{y}$ in an information-reconciliation context. In practical reconciliation schema, $X$ will be encoded with a higher rate than $H(X|Y)$, and the efficiency parameter $f \geq 1$ is defined as:
\begin{equation}
\label{eq:eff1} I_{\textrm{leak}}=fH(X|Y)
\end{equation}
where $I_{\textrm{leak}}$ is the leaked information per bit during the
reconciliation process and $H(X|Y)$ is the conditional Shannon entropy \cite{Elkouss2}.

However, the efficiency is not the only parameter to assess the performance of a reconciliation protocol. Computational complexity and interactivity must also be considered. Low computational complexity makes the reconciliation protocol feasible. Interactivity influences the quality of the protocols in the case of a high latency environment. If the protocol has low interactivity, such as a one-way reconciliation protocol, high latency has no effect on it. By contrast, a high interactivity protocol such as \emph{Cascade} would be greatly affected.

In coding theory, any linear combination of codewords in a linear code is itself also a codeword.
Linear codes have been used for information reconciliation. For
example, Hamming codes were used for \emph{Winnow} \cite{Winnow}
and LDPC codes were used in \cite{{Elkouss1},{Elkouss2},{Elkouss3},{Elkouss4},{Martinez}}.
The code rate of an error-correcting code is the proportion of a data-stream that is meaningful. In other words, if the code rate is
$k/n$, a codeword \textbf{c} of $n$ symbols is composed of $k$
information symbols and $n-k$ redundant symbols.

As defined in (\ref{eq:eff1}), the efficiency of a reconciliation
protocol can be measured by comparing the amount of revealed
information with the conditional Shannon entropy function:
\begin{equation}
f=\frac{I_{\textrm{leak}}}{H(X|Y)}
\end{equation}
Let ``$\textrm{leak}$'' refer to the amount of information correlated with Alice's sequence $X$ that was revealed to Eve during reconciliation. Because $I_{\textrm{leak}}$ is the leaked information per bit and ``$\textrm{leak}$'' is the total amount of leakage, the efficiency parameter can be redefined by \cite{Elkouss4}:
\begin{equation}
f=\frac{\textrm{leak}}{nH(X|Y)}
\end{equation}
As previously mentioned, a quantum channel of QKD is modeled by a BSC with a crossover probability $p$. Then, the correlation of $X$ and $Y$ can be characterized by $p$. Therefore, the reconciliation efficiency is given by:
\begin{equation}
\label{eq:efficiency}
f=\frac{\textrm{leak}}{n\cdot h(p)}
\end{equation}
where $h(\cdot)$ is the binary entropy.

Let $C$ be the set of all possible communication messages during the
reconciliation process. The leaked information of
reconciliation is given by \cite{Renato}:
\begin{equation}
\label{LI} \textrm{leak}:=\log_2|C|-H_{\infty}(C|X)
\end{equation}
where $|\cdot|$ is the cardinality of a set, and
$H_{\infty}(\cdot)$ is the minimum entropy.

Lower efficiency means fewer information leaks during reconciliation. Less information leakage causes the privacy amplification to discard a smaller amount of information. Thus, more information remains in sifted keys, leading to a higher secret key rate. After all, a lower efficiency is directly related to the higher secret key rate, which is why efficiency is a performance indicator of reconciliation protocols.

\subsection{Previous Work}
\label{Subsec:Previous Work}
\emph{Cascade}, proposed by Brassard and Salvail in 1992, was specially designed for information reconciliation \cite{Cascade}. \emph{Cascade} is considered the de-facto standard because of its outstanding performance. \emph{Cascade} is almost ideal because the amount of leaked information can be slightly greater than the theoretical bound when the crossover probability of the quantum channel is less than 15\%. Although the number of channel uses is increased considerably, modified \emph{Cascade} has also been proposed with improved efficiency \cite{Martinez-Mateo}. During real implementation, the error rate of the quantum channel is less than 5\%, making \emph{Cascade} the appropriate protocol for information reconciliation. \emph{Cascade} takes advantage of the interaction between Alice and Bob over an authenticated public channel to simplify the problem of reconciliation. \emph{Cascade} can be described as a very compact and elegant algorithm. The protocol consists of a binary search and trace-back algorithm that iterates over several passes. At each pass, Alice and Bob divide secret keys into several blocks, compute the parities of both sides, and compare them for each block. A parity mismatch indicates an odd number of errors, and binary search allows both parties to find one of the errors. Because a binary search can fix only one error per block, it proceeds to the next pass to correct remaining errors. In the next pass, secret keys with remaining errors are shuffled, and then a binary search is performed again by dividing it into blocks whose sizes are doubled. Alice and Bob perform trace-back after a binary search is completed from pass two or beyond. Error correction in the current pass uncovers errors that were not revealed in the previous pass. These two processes occur in \emph{Cascade} and proceed to the next pass when no additional errors can be found. \emph{Cascade} is terminated at the pass in which errors are no longer found. However, an iterative binary search causes \emph{Cascade} to require too many uses of the  public channel between Alice and Bob.

\emph{Winnow} is another protocol to solve the high interactivity problem of \emph{Cascade} \cite{Winnow}. Like \emph{Cascade}, \emph{Winnow} splits the key strings to be reconciled into blocks. Error correction is based on the Hamming codes instead of a binary search. Alice and Bob exchange the parities of every block. If parity mismatch occurs in some blocks, they exchange a syndrome of Hamming codes, and Bob makes a decision on error position and flips error bits. However, Bob can make a wrong decision when two or more errors occur. Incorrect bit flipping by Bob also creates new error bits. These new errors, called ``introduced additional errors'', are the main reason the efficiency of \emph{Winnow} is lower compared to \emph{Cascade}. Although \emph{Winnow} is considerably faster than \emph{Cascade}, its performance is poor in the error range of interest.

The LDPC forward error correction algorithm was initially proposed by Gallager \cite{LDPC}. It became popular in the early 2000s for digital communications and was first applied to the QKD protocol by BBN \cite{BBN}. The main benefits of protocols using LDPC codes are that only a single round-trip communication is required and the amount of information that may be exposed to an eavesdropper is more easily computed compared to \emph{Cascade}, which requires several round-trip communications. The use of LDPC codes especially optimized for BSC rather than simple applications was first proposed in \cite{Elkouss1}. However, the efficiency curve exhibited a saw behavior as a result of a lack of code rate adaptation. To solve the saw behavior problem, the authors of \cite{Elkouss1} proposed a rate-adaptive scheme \cite{Elkouss2} as well as  ``blind reconciliation" \cite{Elkouss3}, which is slightly more interactive and has smaller block sizes for more realistic hardware implementation.

The use of polar codes under the SC decoder in reconciliation was proposed in \cite{Jouguet}. Jouguet provided performance results for polar codes applied to QKD with code length over $2^{16}$. However, the performance of polar codes under the SC decoder falls behind that of LDPC codes. More specific information on polar codes in QKD has been introduced, and specifically designed polar codes for the QKD environment were suggested in \cite{PolarQKD}.

Both methods utilize the SC decoder in the application to QKD. However, empirical studies show that for short and moderate block sizes, the SC decoding of polar codes in digital communications does not achieve the performance of turbo or LDPC codes. To narrow this gap, Tal and Vardy proposed the SCL decoder \cite{ListDecoding}. In the SCL decoder, $L$ possible codewords are considered concurrently and the most likely codeword is selected. Unfortunately, although SCL decoding effectively improves the performance of polar codes, it still falls short compared to LDPC and turbo codes of comparable length. The authos tried to improve performance by applying CRC precoding to SCL decoding. As a result, the bit error rate (BER) of the SCL decoder with CRC is better than that achieved with LDPC codes currently used in the WiMAX standard.

\subsection{Classical Post-Processing in QKD Experiments}
\label{Subsec:CPP}
Classical post-processing is a procedure that includes information reconciliation and privacy amplification in a classical public channel. Although not mentioned in introduction, a step exists known as error verification between information reconciliation and privacy amplification. Regardless of the scheme used, residual errors occur that Alice and Bob cannot identify. After reconciliation, Bob transmits hash tags of reconciliation blocks to Alice. If the hash tags do not match, the associated block is discarded \cite{experiment2,experiment22}.

Since QKD was proposed in 1984, starting from its early demonstration in feasibility experiments \cite{BBBSS,pp}, faster and longer-distance systems have been proposed. However, few experiments have been conducted that explicitly show how CPP is performed. Recently, three experiments were conducted that achieved excellent results revealing how CPP is performed.
In the first experiment \cite{experiment1}, the input size of privacy amplification was $6.6\times10^5$ and \emph{Cascade} was used for reconciliation. Although it was not clearly specified, the reconciliation block size of \emph{Cascade} was less than $6.6\times10^3$.
In the second \cite{experiment2} and third experiment \cite{experiment22}, the the input size of privacy amplification was $10^6$ and LDPC codes of 1944 bits were used.

As shown in the examples of the three experiments, the size of the reconciliation block was less than $6.6\times10^3$. In the second and third experiments, used linear block codes were only 1944 bits long. This is why we are interested in reconciliation using short-length polar codes. Reconciliation using short-length LDPC codes was already proposed in \cite{Elkouss3}. The results of a CRC-aided SCL decoder of polar codes showed that polar codes of length $2048$ outperformed LDPC codes of similar length \cite{ListDecoding}. Therefore, the performance of small block reconciliation can be improved using polar codes under a CRC-aided SCL decoder.

\section{Improved Reconciliation with Polar Codes}
\label{Sec:Imroved Reconciliation}
As mentioned in Section~\ref{Subsec:Previous Work}, rate-adaptive LDPC codes with small block size are proposed for QKD hardware implementation.
In addition to their excellent performance in digital communications, the use of polar codes under a CRC-aided SCL decoder can improve the performance of reconciliation. In this study, we propose a reconciliation protocol for real hardware implementations utilizing a CRC-aided SCL decoder of polar codes.

\subsection{Polar Codes}
\label{Subsec:Polar Codes} 
The development of polar codes by  Arikan \cite{PolarCodes} was a breakthrough in coding theory. Polar codes have been proven to achieve the capacity of symmetric binary-input discrete memoryless channels (BI-DMC) with an explicit construction. Moreover, polar codes have low encoding and decoding complexity, which makes them fit for efficient hardware implementation \cite{PC_HW}.


Let the code under consideration have length $n=2^m$ for some $m\geq0$ and dimension $k$; thus, the number of redundant bits is $n-k$. 
We use the notations $\textbf{a} = (a_i)_{i=0}^{n-1} = a_0^{n-1}$ as a row vector. A codeword \textbf{x} is generated by
\begin{equation}
\label{eq:encoding}
\textbf{x} = \textbf{u}G 
\end{equation}
where $G$ is the generator matrix and \textbf{u} is a input vector. 
For $\mathcal{A}$ a subset of $\{0, ..., n-1\}$, (\ref{eq:encoding}) may be written as:
\begin{equation}
\label{eq:encoding2}
\textbf{x} = \textbf{u}_{\mathcal{A}}G(\mathcal{A})\oplus\textbf{u}_{\mathcal{A}^c}G(\mathcal{A}^c	)
\end{equation}
where $G(\mathcal{A})$ denotes the submatrix of $G$ formed by the rows with indices in $\mathcal{A}$. A set $\mathcal{A}$ is the information set. The information set is determined by calculating the rate of polarization of individual channels, which is called the Bhattacharyya parameter:
\begin{equation}
Z(W_n^{(i)})=\sum_{y_0^{n-1}\in\mathcal{Y}^{N}}\sum_{u_0^{i-1}\in\mathcal{X}^{i}}\sqrt{P_W^{(i)}(\textbf{y}, {\hat{u}}_0^{i-1}|0) P_W^{(i)}(\textbf{y}, {\hat{u}}_0^{i-1}|1)}
\end{equation}
where $Z(W_n^{(i)})$ is an upper bound on the probability of maximum-likelihood (ML) decision error when $W_n^{(i)}$ is used once to transmit a 0 or 1. The information set $\mathcal{A}$ is chosen as the $k$-element subset of $\lbrace 0,...,n-1 \rbrace$ such that $Z(W_n^{(i)})\leq Z(W_n^{(j)})$ for all $i\in\mathcal{A}, j\in\mathcal{A}^c$. We also call $\textbf{u}_{\mathcal{A}^c}$ as frozen bits or vector.

After the codeword is sent over the channel, Bob obtains the vector $\textbf{y} = y_0^{n-1}$. The SC decoder is applied to \textbf{y} given frozen bits, which results in an estimated codeword $\hat{\textbf{x}}$ or information vector $\hat{\textbf{u}}$.

The SC decoder consists of $n$ decision elements (DEs), one for each information element $u_i$. The $i$th DE behaves like the binary-input coordinate channel \[W_n^{(i)} : \mathcal{X}\rightarrow\mathcal{Y}^N\times\mathcal{X}^{i-1}, 0\leq i\leq n-1\] defined by the transition probabilities
\begin{equation}
P_W^{(i)}(\textbf{y}, {\hat{u}}_0^{i-1}|u_i)=\sum_{u_{i+1}^n\in\mathcal{X}^{n-i}}
\frac{1}{2^{N-1}} P_W(\textbf{y}|\textbf{u})
\end{equation}
where $(\textbf{y}, {\hat{u}}_0^{i-1})$ denotes the output of $W_n^{(i)}$ and $u_i$ is its input.

The DEs observe $(\textbf{y}, u_0^{i-1})$ and make decision $\hat{\textbf{u}}$ of ${\textbf{u}}$ by computing the likelihood ratio (LR):
\begin{equation}
L_n^{(i)}(\textbf{y},u_0^{i-1})=\frac{P_W^{(i)}(\textbf{y}, {\hat{u}}_0^{i-1}|0)}
{P_W^{(i)}(\textbf{y}, {\hat{u}}_0^{i-1}|1)}
\end{equation}
in the order 0 to $n-1$, and generate its decision as:
\[\hat{u}_i=
\begin{cases}
u_i, & \text{ if } i \in \mathcal{A}^c \\ 
0, & \text{ if } i \in \mathcal{A} \text{ and } L_n^{(i)}(\textbf{y},u_0^{i-1})\geq 1  \\ 
1, & \text{ otherwise }
\end{cases}\]
which is sent to all subsequent DEs.

\subsection{Source Polarization}
\label{Subsec:Source Polarization}
Information reconciliation can be substituted for source coding with side information. Therefore, we approach this problem from the point of view of source coding using polar codes, a method that has already proposed and organized \cite{SP}. The notion ``source polarization" complements ``channel polarization" that was studied in \cite{PolarCodes}. Lossless source coding with side information using polar codes can be performed using the analysis of source polarization.

In source polarization, bits in a high-entropy (index) set $E_{X|Y}$ plays the role of frozen bits of channel polarization. Let $\alpha$ be the size of a high-entropy set. A high-entropy set indicates $\alpha$ sub-channel indices of the largest conditional entropy term $\lbrace H(U_i|Y^N,U^{i-1})\rbrace_{i=0}^{n-1}$, where $X_i$ denotes the $i$th drawing from $X$ independently, and $X^n$ denotes the $n$ elements of this sequence. The source Bhattacharyya parameter is defined as:	
\begin{equation}
Z(X|Y)=2\sum_y P_Y(y)\sqrt{P_{X|Y}(0|y)P_{X|Y}(1|y)}
\end{equation}
The source Bhattacharyya parameter satisfies a simple two-by-two polarization transformation, making it easier to determine than a conditional entropy term. Furthermore, $H(X|Y)$ approximates 0 or 1 if and only if $Z(X|Y)$ approximates 0 or 1. Thus, the parameters $\lbrace H(U_i|Y^n,U^{i-1})\rbrace_{i=0}^{n-1}$ and $\lbrace Z(U_i|Y^n,U^{i-1})\rbrace_{i=0}^{n-1}$ polarize simultaneously. We can use the Bhattacharyya parameter instead of conditional entropy to define a high-entropy set $E_{X|Y}$ that is a chosen $\alpha$-element subset of $\lbrace 0,...,n-1 \rbrace$ such that $Z(U_i|Y^N,U^{i-1})>Z(U_j|Y^N,U^{j-1})$ for all $i\in E_{X|Y}, j\notin E_{X|Y}$. 

For simplicity of deployment, we denote $\lbrace u_i |i\in E_{X|Y} \rbrace$ by high-entropy bits, $E^c_{X|Y}$ by the information (index) set, and $\lbrace u_i |i\in E^c_{X|Y} \rbrace$ by information bits just as in channel polarization. Then the code length $n$ will be $\alpha+k$, where $k$ is the size of the information set as in the case of channel polarization.

\subsection{Protocol}
\label{Subsec:Protocol}

We first introduce a reconciliation protocol using polar codes under the SC decoder. For the sake of brevity, we call it the \textsf{SC} protocol after the name of the decoder. The process of the protocol is not much different from that of source polarization. It is also similar to reconciliation using LDPC codes. 

\begin{figure} [htbp]
\centerline{\epsfig{file=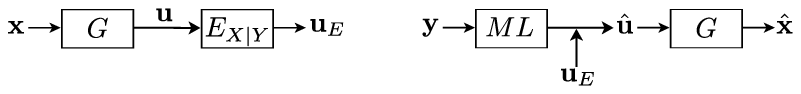, width=8.2cm}} 
\vspace*{13pt}
\fcaption{\label{fig:SC}Procedure of the \textsf{SC} protocol. Given sifted key \textbf{x}, Alice generates and sends high-entropy bits $\textbf{u}_E$(left). Bob performs the SC decoding using sifted key \textbf{y} and high-entropy bits $\textbf{u}_E$(right).}
\end{figure}

Alice and Bob prepare shared information about a high-entropy set of codes. In other words, Bob knows only the positions of high-entropy bits and does not know the values before reconciliation begins. To reconcile $n$-length sifted keys \textbf{x} and \textbf{y}, Alice generates a codeword $\textbf{u}=\textbf{x}G$ and sends only high-entropy bits $\textbf{u}_E$ that are a part of \textbf{u}. When Bob receives them, he knows all the positions and values of high-entropy bits. On Bob's side, the SC decoder makes decisions $\hat{\textbf{u}}$, treating high-entropy bits as frozen bits. Therefore, the leaked information of the \textsf{SC} protocol represents information about high-entropy bits. The procedure of the \textsf{SC} protocol is shown in Fig. \ref{fig:SC}.

Sifted keys cannot be CRC-precoded because they are obtained arbitrary. For CRC precoding, we introduce a method inspired by \cite{PolarQKD} that exploits another arbitrary string. We call it a virtual string because it is not actually transmitted. We suggest an information reconciliation protocol using polar codes with a CRC-aided SCL decoder with a virtual string. We call this protocol the \textsf{CL} protocol, named after a CRC-aided SCL decoder. The code length $n$ will be $\alpha+k+c$ where $c$ is the size of a CRC.
The procedure of the protocol is shown in Fig. \ref{fig:CL} and described as follows:


\begin{figure} [tbp]
\centerline{\epsfig{file=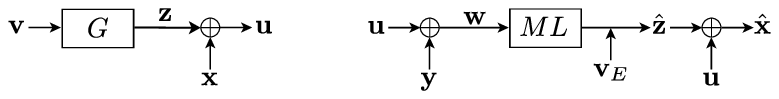, width=8.2cm}} 
\vspace*{13pt}
\fcaption{\label{fig:CL}Procedure of the \textsf{CL} protocol. Alice generates a codeword \textbf{u} using virtual string \textbf{v} and sifted key \textbf{x}(left). Estimated key $\hat{\textbf{x}}$ is produced from the result of a decoder $\hat{\textbf{z}}$ and a codeword \textbf{u}(right).}
\end{figure}

\begin{romanlist}
 \item \textit{Setup}: Fix the length of strings $n=2^m$ for some $m\geq 1$ and the size of a high-entropy set $\alpha=f\cdot n\cdot H(X|Y)$. The rate $R$ of the polar code is $k/n$,  High-entropy bits $\textbf{v}_E$ are all zero and available to both Alice and Bob. Let sifted keys $\textbf{x}$ and $\textbf{y}$ be $n-$length strings, which are instances of random variables $X$ and $Y$, respectively, where $X$ and $Y$ are the input and output, respectively, of a BSC with a crossover probability $p$. 
 \item \textit{Precoding}: Alice prepares an $n-$length virtual string $\textbf{v}$ that is composed of $\alpha$ zero bits in positions of a high-entropy set and $k$ random bits in positions of an information set. She computes a CRC of first $\alpha+k$ bits of $\textbf{v}$ and fills the remaining $c$ positions of $\textbf{v}$ with CRC bits. Therefore, a virtual string consists of $\alpha$ high-entropy bits, $k$ random bits, and $c$ CRC bits.
 
 \item \textit{Encoding}: Alice generates $n-$length string $\textbf{z}=\textbf{v}G$ by the encoder of polar codes. Alice calculates a codeword $\textbf{u}=\textbf{z}\oplus\textbf{x}$ and sends it through an error-free public channel.
 
 \item \textit{Decoding}: Bob constructs a string $\textbf{w}=\textbf{u}\oplus\textbf{y}$. Given $\textbf{w}$ and a high-entropy set, the decoder sequentially generates an estimate $\hat{\textbf{z}}$ of $\textbf{z}$ by the rule:
\[\hat{z}_i=
\begin{cases}
z_i, & \text{ if } i \in E_{X|Y} \\ 
0, & \text{ if } i \notin E_{X|Y} \text{ and } L_n^{(i)}(\textbf{z},\hat{v}_0^{i-1})\geq 1  \\ 
1, & \text{ otherwise }
\end{cases}\]
where
\[L_n^{(i)}(\textbf{z},\hat{v}_0^{i-1})=\frac{P_{V|Z^N V^{i-1}}(0|\textbf{z}, {\hat{v}}_0^{i-1})}
{P_{V|Z^N V^{i-1}}(1|\textbf{z}, {\hat{v}}_0^{i-1})}\]
is the likelihood ratio in source polarization. Bob chooses the most probable estimations among the list that have a correct CRC. Therefore, Bob recovers $\hat{\textbf{x}}=\hat{\textbf{z}}\oplus\textbf{u}$, and if $\hat{\textbf{x}}=\textbf{x}$, the protocol completed successfully.
\end{romanlist}

\subsection{Leaked Information}
\label{Subsec:Leaked Information}
Reconciliation using linear codes uses $n-k$ redundant symbols to reconcile Alice's and Bob's strings. In the case of protocols using LDPC codes, exact $n-k$ syndrome bits are sent over a public channel. Furthermore, the \textsf{SC} protocol uses $n-k$ high-entropy bits to reconcile. The amount of leakage of protocols using LDPC codes or polar codes can be calculated as the number of redundant symbols. The leaked information of reconciliation can be calculated as previously shown in (\ref{LI}).

\vspace*{12pt}
\noindent
{\bf Corollary~1:} Let \rm\textsf{SC} be the information reconciliation protocol using polar codes with the SC decoder defined in Section~\ref{Subsec:Protocol}. Then, the leakage is given by:
\[\rm{leak}_{\mathsf{SC}} \it=n-k\]
where n is the length of codewords and $k$ is the size of the information set.

\vspace*{12pt}
\noindent
{\bf Proof:} In the \textsf{SC} protocol, communication messages are high-entropy bits $\textbf{u}_E$. Therefore, leaked information of the \textsf{SC} protocol is computed by
\[\textrm{leak}_{\textsf{SC}}=\log_2{|U_E|}-H_\infty(U_E|X)\]
where $U_E$ denotes high-entropy bits of polar codes and $\log_2{|U_E|}$ refers to the number of communication message bits that is $n-k$. The term
$H_\infty(U_E|X)$ is computed as:
\[H_\infty (U_E|X)=\min_{\textbf{x}} H_\infty (U_E|\textbf{x})=\min_{\textbf{x}} \lbrace -\log_2 \max_{\textbf{u}_E} P_{U_E|X}(\textbf{u}_E|\textbf{x}) \rbrace \]

When $\textbf{x}$ is measured, $\textbf{u}_E$ is determined to be a part of $\textbf{u}=\textbf{x}G$ and $H_\infty (U_E|X)$ is zero. Therefore, total leakage is $n-k$ \square\,.


\vspace*{12pt}
\noindent
{\bf Corollary~2:} Let \textsf{CL} be an information reconciliation protocol using polar codes with a CRC-aided SCL decoder defined in Section~\ref{Subsec:Protocol}. Then, the leakage is given by:
\[\rm{leak}_{\mathsf{CL}} \it=n-k\]
where $n$ is the length of codewords, and $k$ is the size of the information set.

\vspace*{12pt}
\noindent
{\bf Proof:} In the \textsf{CL} protocol, communication messages are codewords $\textbf{u}$. We then compute leakage of the \textsf{CL} protocol by 
\[\textrm{leak}_{\textsf{CL}}=\log_2{|U|}-H_\infty(U|X)\]
where $U$ is the set of all possible $\textbf{u}$. Codeword $\textbf{u}$ is calculated by $\textbf{z}\oplus\textbf{x}$ and $n$-length string $\textbf{x}$ is an instance of a random variable $X$. Therefore $\log_2|U|$ is $n$.

 The term $H_\infty(U|X)$ is calculated as:
\[H_\infty (U|X)=\min_{\textbf{x}} H_\infty (U|\textbf{x})=\min_{\textbf{x}} \lbrace -\log_2 \max_{\textbf{u}} P_{U|X}(\textbf{u}|\textbf{x}) \rbrace\]

We consider inside term $\max_{\textbf{u}} P_{U|X}(\textbf{u}|\textbf{x})$ first.  
A virtual string \textbf{v} includes only $k$ random bits that are equally probable. Consequently, every possible $\textbf{z}=\textbf{v}G$ is equiprobable and the number of possible cases of the string $\textbf{z}$ is $2^{k}$. Codeword $\textbf{u}=\textbf{z}\oplus\textbf{x}$ is equiprobable given a measurement \textbf{x}. 

 Every $P_{U|X}(\textbf{u}|\textbf{x})$ is the same regardless of what \textbf{u} is:
\[\min_{\textbf{x}} \lbrace -\log_2 \max_{\textbf{u}} P_{U|X}(\textbf{u}|\textbf{x}) \rbrace = \min_{\textbf{x}} \lbrace -\log_2 P_{U|X}(\textbf{u}|\textbf{x}) \rbrace\]

Then, the minimum entropy can be calculated as:
\[\min_{\textbf{x}} \lbrace -\log_2 P_{U|X}(\textbf{u}|\textbf{x}) \rbrace =\min_{\textbf{x}}\lbrace -\log_2 2^{-k}\rbrace =k\]
We can conclude that the leaked information of reconciliation is given by
$n-k$  \square\,.

Because the proposed protocol transmits $n$ bits over the public channel, the amount of leaked information appears to be larger than other protocols. We showed that the leakage of the proposed protocol is  $n-k$ the same as other protocols through Corollary 1 and 2.

\section{Results}
\label{Sec:Results}

In this section, we discuss the efficiency of the proposed protocol by comparing the results of protocols using LDPC codes. Through these simulations, we showed that the proposed protocol experiences smaller leakage compared to protocols using LDPC codes for several quantum BER (QBER) over the BSC. The block size is crucial, as it affects the secret key generation rate of QKD. Because the explicit reconciliation block size of actual QKD experiments is less than $6.6\times10^{3}$, we performed the simulations with limited block size. We also considered the complexity of protocols. 

\subsection{Simulation Results}
\label{Subsec:Simulation Results}

\begin{figure} [htbp]
\centerline{\epsfig{file=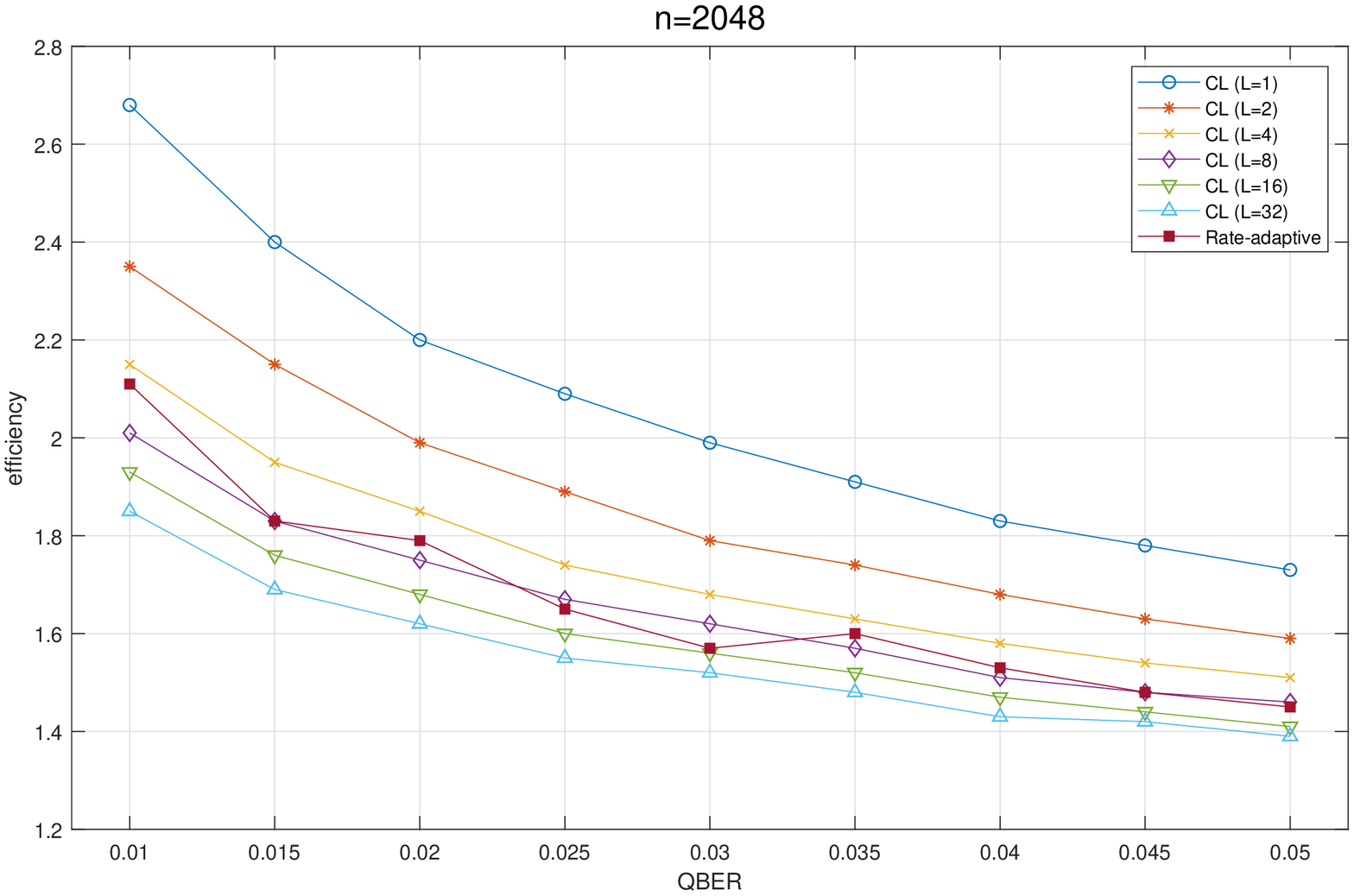, width=12cm}} 
\centerline{\epsfig{file=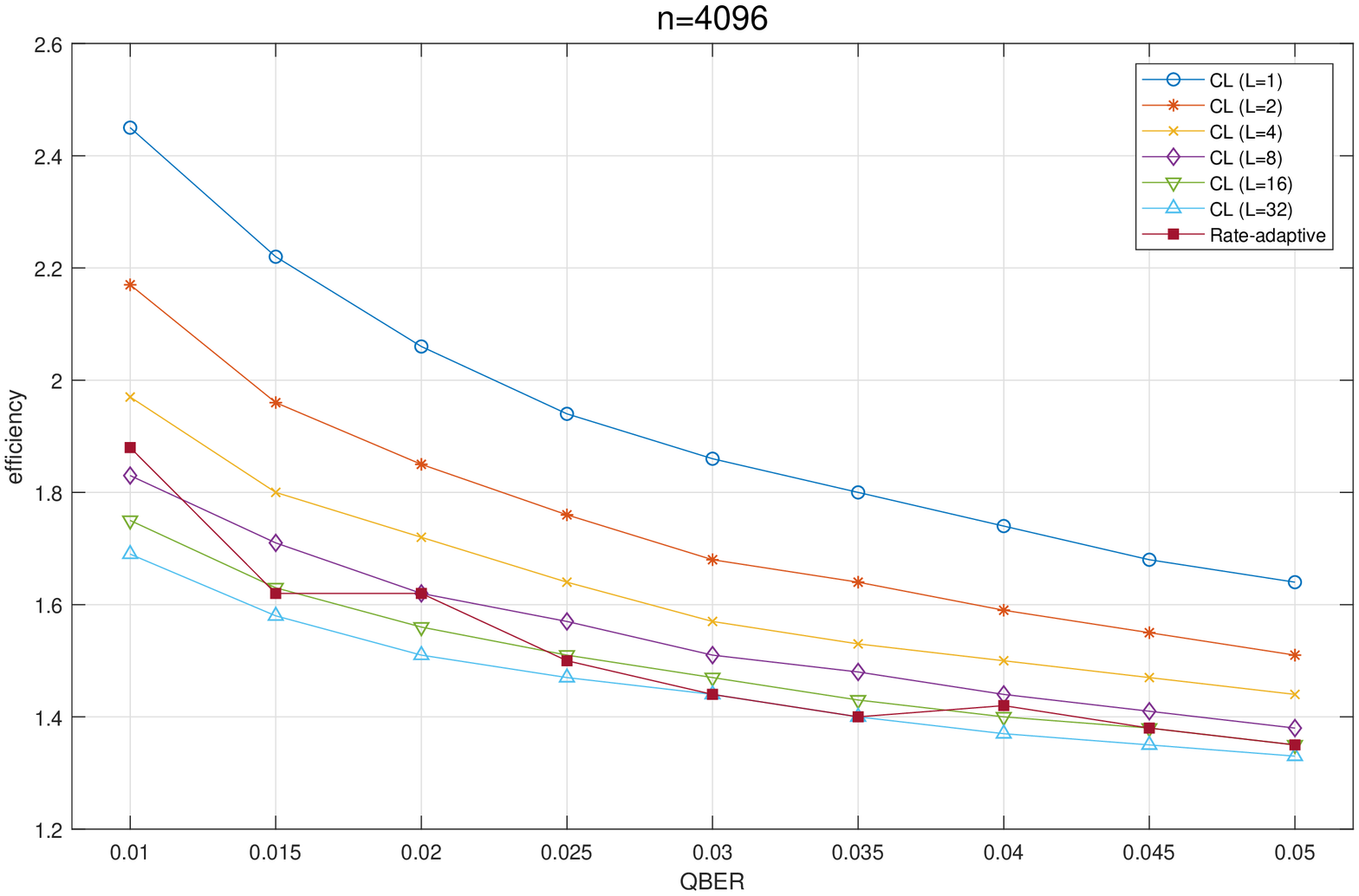, width=12cm}} 
\vspace*{13pt}
\fcaption{\label{fig:eff}Performance of the rate-adaptive and \textsf{CL} protocols with block size of $2048$(top) and $4096$(bottom). At block size of $2048$(top), LDPC codes has following rates: 0.8 for QBER of 0.010, 0.7 for QBER range [0.015, 0.030] and 0.6 for QBER range [0.035, 0.050]. At a block size of $4096$(bottom), LDPC codes has following rates: 0.8 for QBER range [0.010, 0.015], 0.7 for QBER range [0.020, 0.035] and 0.6 for QBER range [0.040, 0.050]. Polar codes have no specific mother rate; list size is $L=1, 2, 4, 8, 16$ and $32$, and CRC is 16 bits long. The result of list size 1 is the performance of a protocol using polar codes under an SC decoder.}
\end{figure}

Fig. \ref{fig:eff} shows the efficiency, defined in  (\ref{eq:efficiency}), of the proposed \textsf{CL} protocol as well as the rate-adaptive protocol proposed in \cite{Elkouss2}. 
The efficiency of the \textsf{CL} protocol is calculated by \eqref{eq:efficiency} using Corollary~2.
The simulations were conducted with block sizes of $2048$ and $4096$. In each simulation, the efficiency was divided by a unit of 0.01, and the simulation points that most approximated the frame error rate (FER) of $10^{-3}$ were selected. The efficiency points selected are shown in curves revealing FER as a function of efficiency for two block sizes in the QBER range $[0.010, 0.050]$, which is Fig. \ref{fig:A1}-\ref{fig:A3} in Appendix A.	

Code construction of polar codes in each QBER was performed using the Monte-Carlo method \cite{PolarCodes}. We constructed codes differently for each QBER to obtain better results. However, a single code designed for specific QBER can be used for multiple QBERs. We computed and sorted the source Bhattacharyya parameter $Z(U_i|Y^N,U^{i-1})$ of each symbol. The size of the high-entropy set was computed as $\lceil f\cdot n\cdot h(p) \rceil$. The elements of a high-entropy set were determined by choosing channels that had the largest $\lceil f\cdot n\cdot h(p) \rceil$ source Bhattacharyya parameters.

LDPC codes were constructed using generator polynomials in \cite{Elkouss3}. Simulations were computed using three coding rates: $R=0.6, 0.7$, and $0.8$. LDPC codes with different code rates were used depending on the QBER. The ratio of shortening and puncturing of the rate-adaptive protocol was fixed to 10\% for all QBERs. An LDPC decoder based on the syndrome decoding algorithm \cite{SyndromeDEC} with a maximum of 200 iterations $(I_{max}=200)$ was utilized.

The top panel of Fig. \ref{fig:eff} shows that the \textsf{CL} protocol with list sizes of 16 or bigger outperformed the rate-adaptive protocol in all QBER ranges. The \textsf{CL} protocol with a list size of 8 also performed better in some QBER ranges. The bottom panel of Fig. \ref{fig:eff} shows that at a block size of $4096$, the proposed protocol still had better reconciliation efficiency than did the rate-adaptive protocol. 

\subsection{Complexity}
\label{Subsec:Complexity}

\noindent
The complexity is a crucial issue in hardware implementation. In general, better efficiency guarantee faster speed in a QKD environment. However, if the complexity of a protocol increases exponentially, using it in experiments will be difficult. Therefore, we computed the complexity of the \textsf{CL} and rate-adaptive protocols. We compared consumption times for decoding of LDPC and polar codes. Decoding of both codes was performed as parallel as possible. All simulations were run on a single core of an Intel i7-4790 CPU with a base clock frequency of 3.60 GHz and maximum turbo frequency of 4 GHz. The block size was $2048$ bits and the QBER range was $[0.01, 0.05]$. Table \ref{table:complexity} shows the summary of the calculation of time complexity.

\begin{table}[htbp]
\tcaption{Complexity of the rate-adaptive (LDPC) and the \textsf{CL} (polar) protocols}
\centering
    \begin{tabular}{ c | c | c }
    \hline
     & \textbf{LDPC} & \textbf{Polar} \\ \hline
    \textbf{Unit}& iteration &  time slot \\ \hline
    \textbf{Number of units}& 200 &  4094 \\ \hline
    \textbf{Average time of each unit}& 1.23 $\mu$s &  0.0615 $\mu$s \\ \hline
    \textbf{Total time}& 246 $\mu$s &  251.78 $\mu$s \\ \hline
    \end{tabular}
\label{table:complexity}
\end{table}

In the case of a fully parallel LDPC decoder, the time unit is an iteration. An average of 1.23 $\mu$s is spent at each iteration. The standard number of iterations was 200. Therefore, we set the number of iterations to 200, and the total time of the protocol was 246 $\mu$s.

A fully parallel implementation for a polar decoder was introduced by Arikan \cite{PolarCodes}. In that work, the time unit of the implementation was a time slot. In our simulations, the average of time slots was 0.0615 $\mu$s. Fully parallel implementation of the polar decoder had a latency of $2n-2$ time slot for a code of block size $n$. Consequently, total time was 251.78 $\mu$s.
From the results, total times of the rate-adaptive and \textsf{CL} protocols were nearly the same in a block size 2048. 

The total time is a product of the average time of each unit and the number of units. Each unit of fully parallel implemented LDPC decoding is composed of check and variable node operations. 
Thus, a block size barely affect calculation times of operations. Every result of the rate-adaptive protocol was obtained with maximum iteration 200 which is the number of units. Consequently, the average time of the rate-adaptive protocol is almost constant regardless of a block size. 

The average time of each unit of polar decoding is always constant regardless of a block size. However, the number of units of polar codes are $2n-2$ which is a function of a block size. Therefore, total time of the proposed protocol increases linearly with a block size.

The total time of LDPC decoding is constant and that of polar decoding is linear function of a block size $n$.
In a block size $4096$, the efficiency of the proposed protocol is still better but twice more time-consuming than that of the rate-adaptive protocol. It can be trade-off between the efficiency and the implementation time. By contrast, in block size $1024$, the proposed protocol will have better efficiency and twice less time-consuming.

The results of efficiency and complexity show that the \textsf{CL} protocol is ideal at a block size of $2048$ or less, as it performs better and the consumption time is nearly the same or less. Depending on experimental environment, the \textsf{CL} protocol may also be better at a block size of $4096$. The performance of the proposed protocol remains better at a block size of $4096$, but shows a longer implementation time. 

We used only an early implementation introduced by Arikan. However, much research has been conducted to reduce the complexity of a polar decoder \cite{SSC, SSC2, SSC3}. Using these studies, the total time of the proposed protocol can be decreased.

The proposed protocol has the advantages of using only a single round-trip communication in reconciliation and incurring no penalty in hardware implementation.

\section{Conclusion}
\label{Sec:Conclusions}
We proposed a reconciliation protocol using polar codes for a QKD system. The proposed protocol guarantees good performance in low QBER and small block size. 

Our simulation results show that the proposed protocol has lower leaked information and reasonable complexity compared to the protocol using LDPC codes. 
We showed that the proposed protocol is ideal for experiments. In experiments with block size of $2048$ or smaller, the proposed protocol showed excellent efficiency and low time complexity. The performance of the proposed protocol was still better at block size of $4096$, but had a longer implementation time. 

The proposed protocol improves on \emph{Cascade}, allowing high efficiency. It also has the practical advantage of low interactivity. The high efficiency derives from the fact that the performance of polar codes with a small block size is superior to that of LDPC codes.


\nonumsection{Acknowledgments}
\noindent
This work was supported by the ICT R\&D program of MSIP/IITP [1711028311, Reliable crypto-system standards and core technology development for secure quantum key distribution network].

\nonumsection{References}

\appendix
\noindent
In this appendix, we present curves showing FER as a function of efficiency for two block sizes in the QBER range $[0.010, 0.050]$. These curves show that FER changes depend on efficiency and those points are selected in Fig. \ref{fig:eff}.

\begin{figure} [h]
\centerline{\epsfig{file=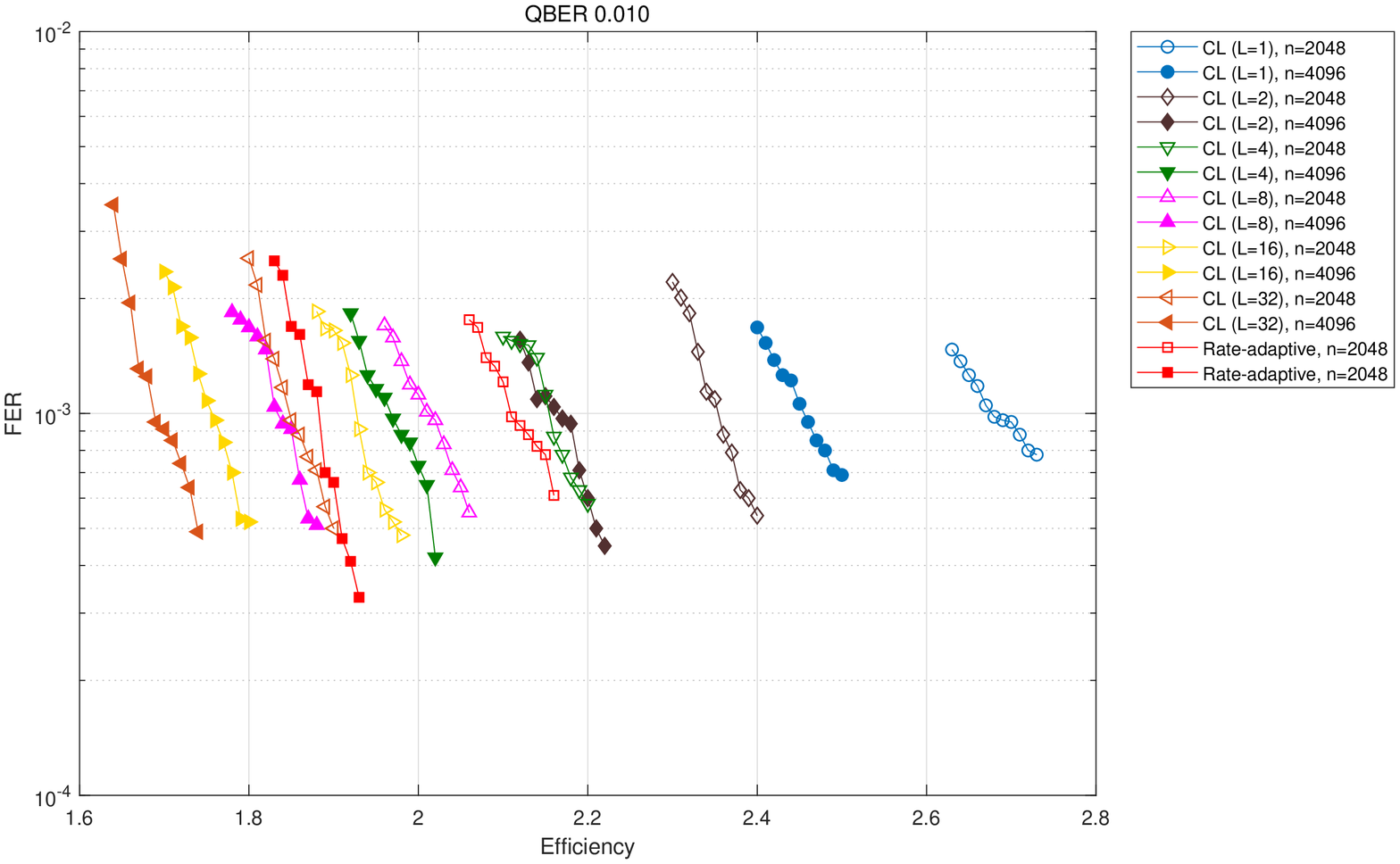, width=11cm}} 
\centerline{\epsfig{file=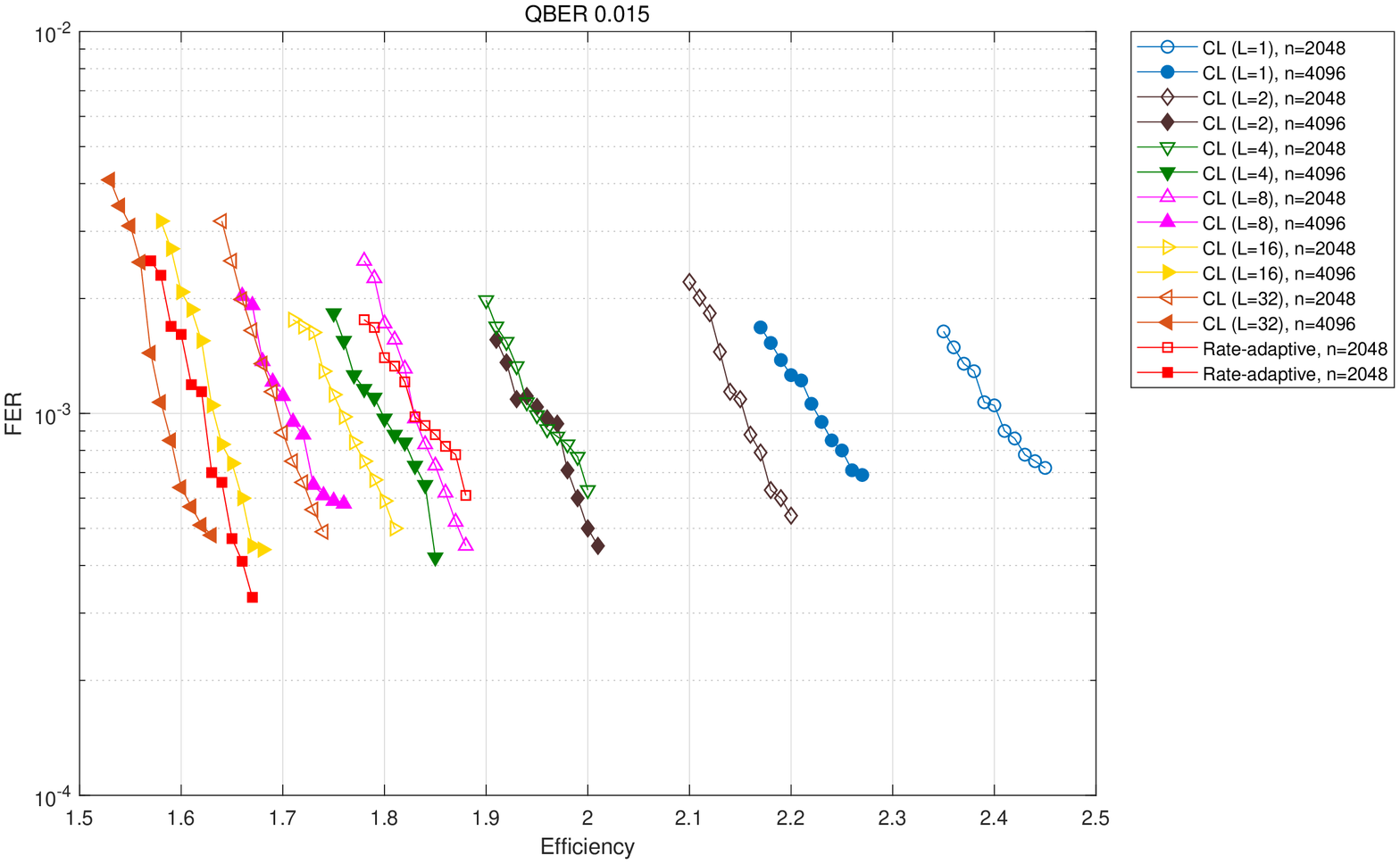, width=11cm}} 
\centerline{\epsfig{file=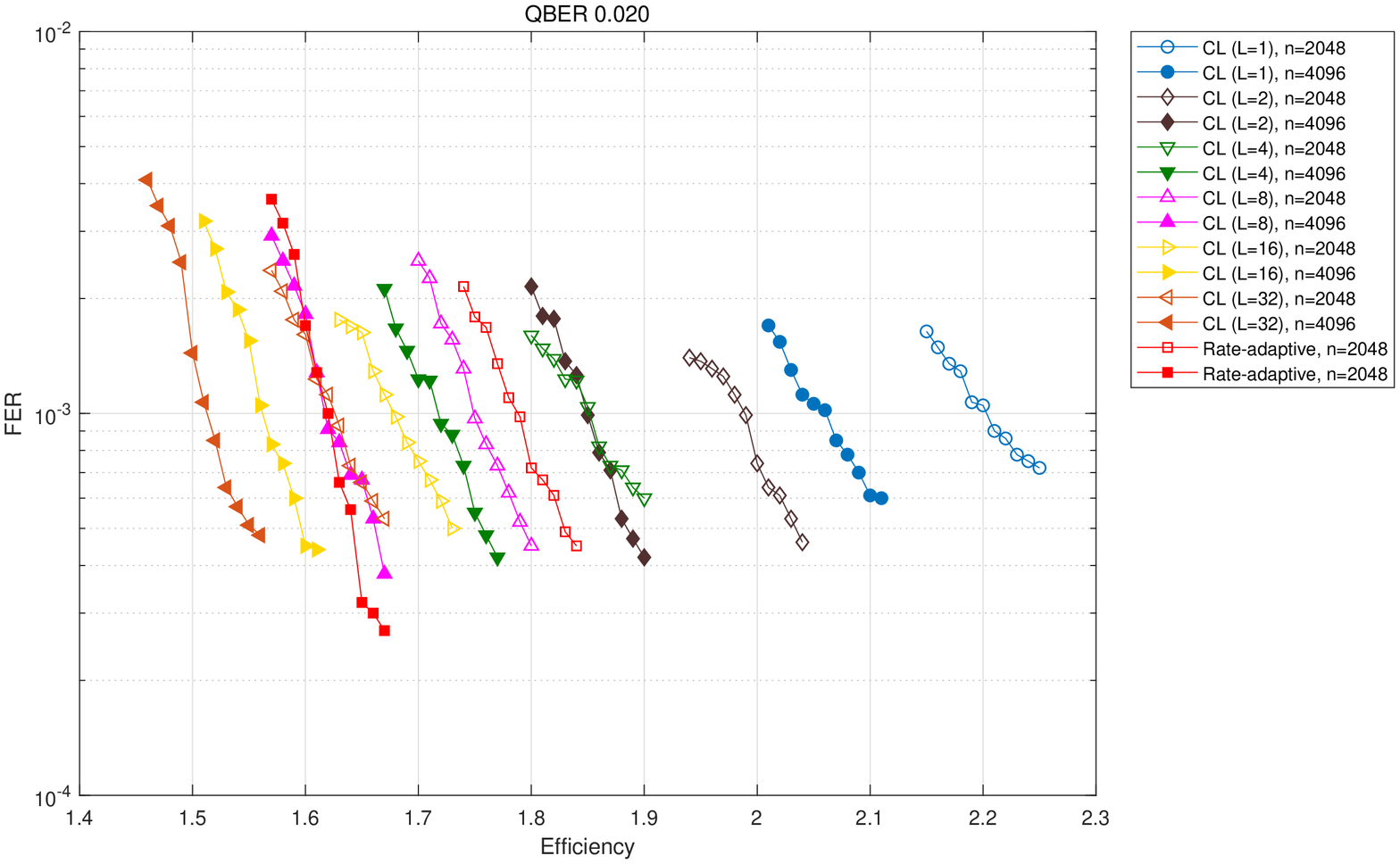, width=11cm}} 
\vspace*{13pt}
\fcaption{\label{fig:A1}FER curves in QBER of 0.010(top), 0.015(middle) and 0.020(bottom)}
\end{figure}

\begin{figure} [h]
\centerline{\epsfig{file=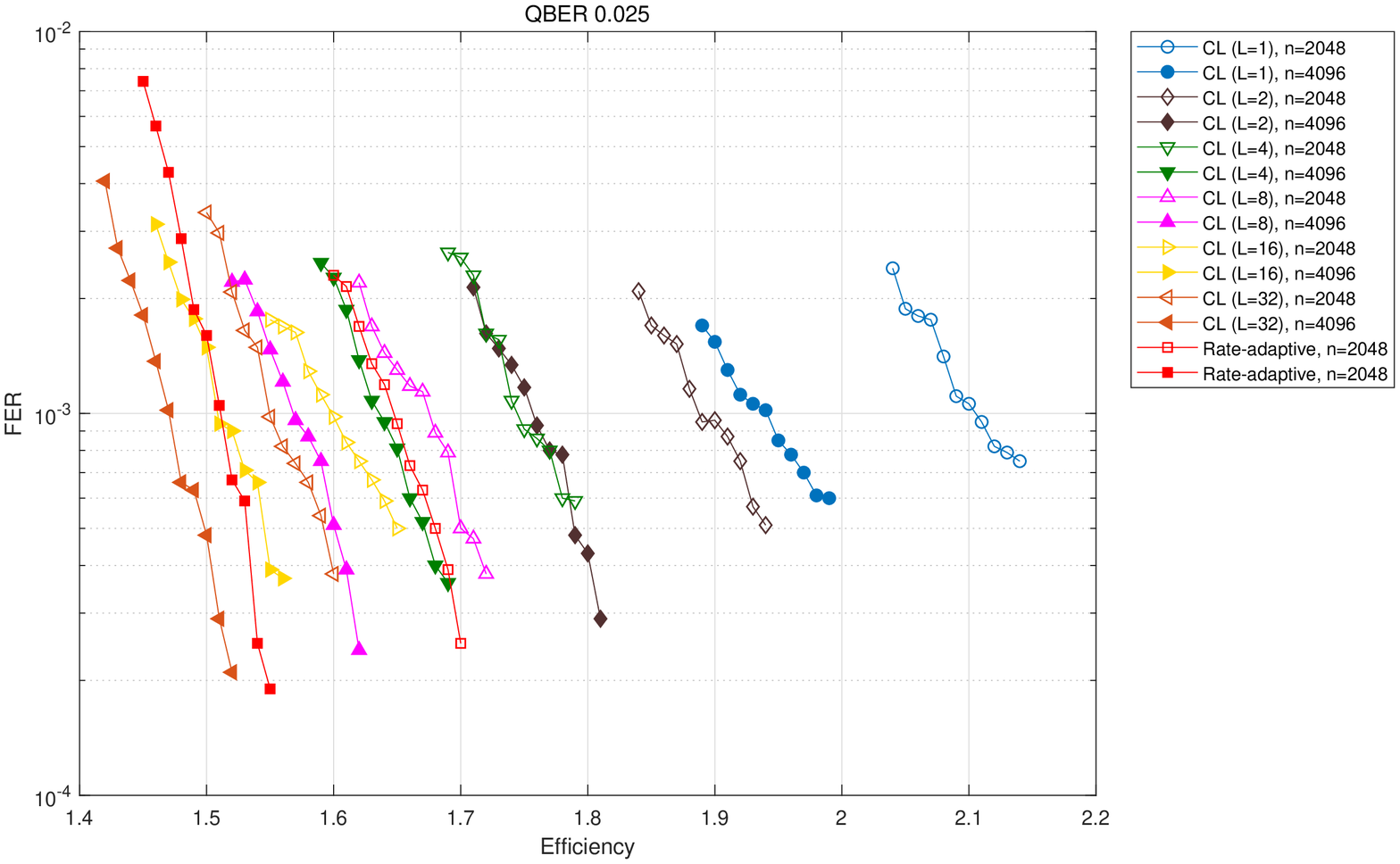, width=11cm}} 
\centerline{\epsfig{file=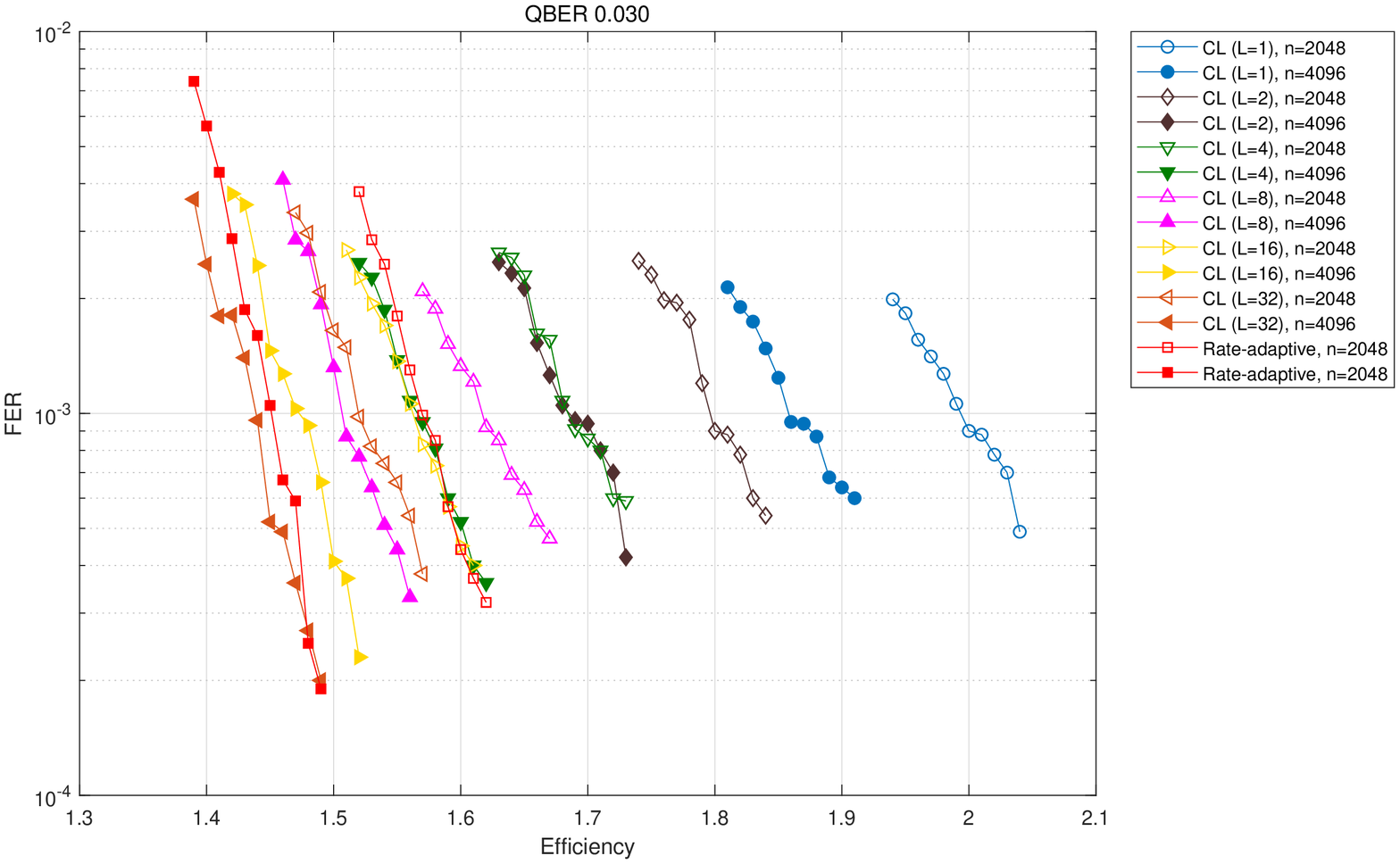, width=11cm}} 
\centerline{\epsfig{file=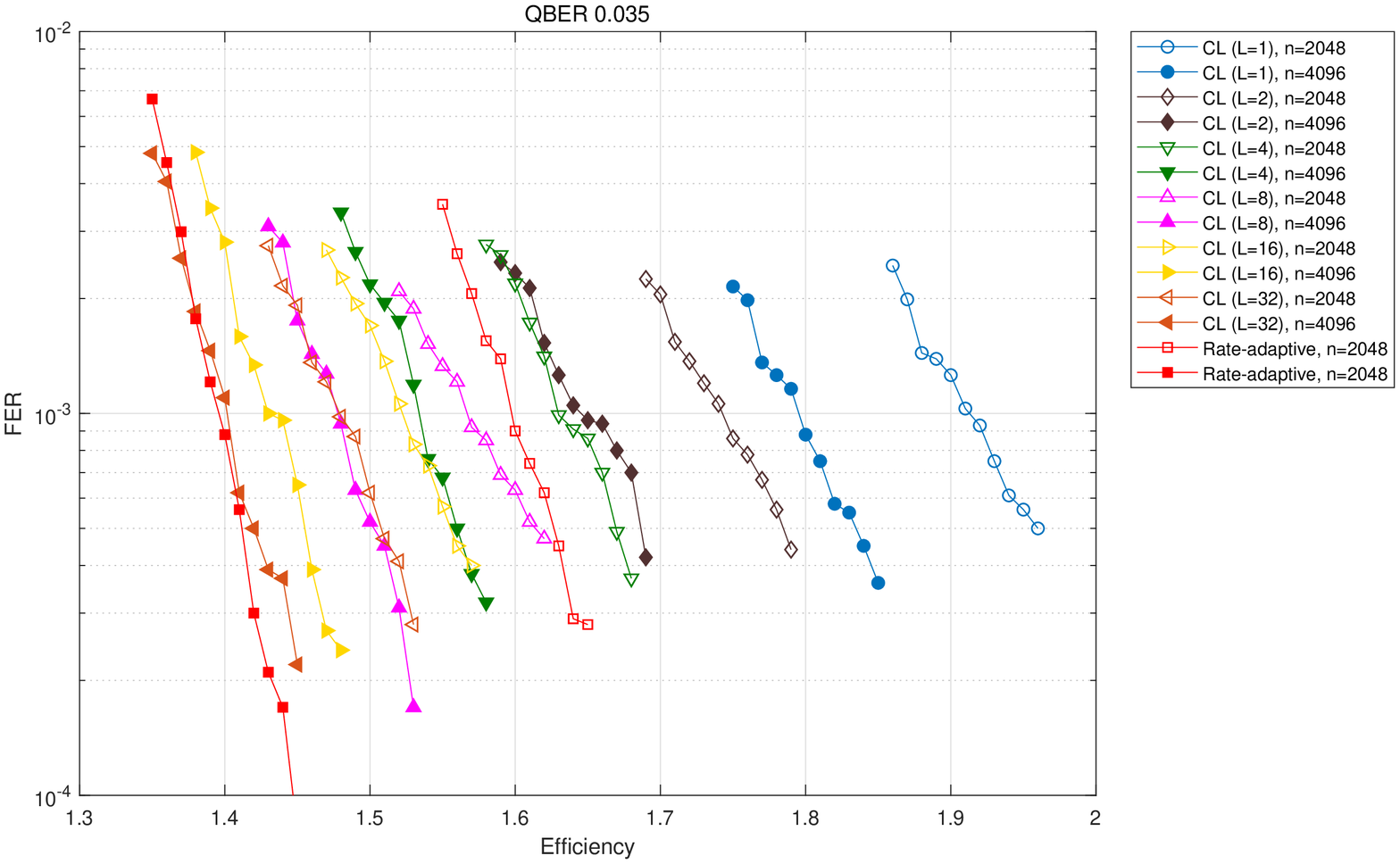, width=11cm}} 
\vspace*{13pt}
\fcaption{\label{fig:A2}FER curves in QBER of 0.025(top), 0.030(middle) and 0.035(bottom)}
\end{figure}

\begin{figure} [h]
\centerline{\epsfig{file=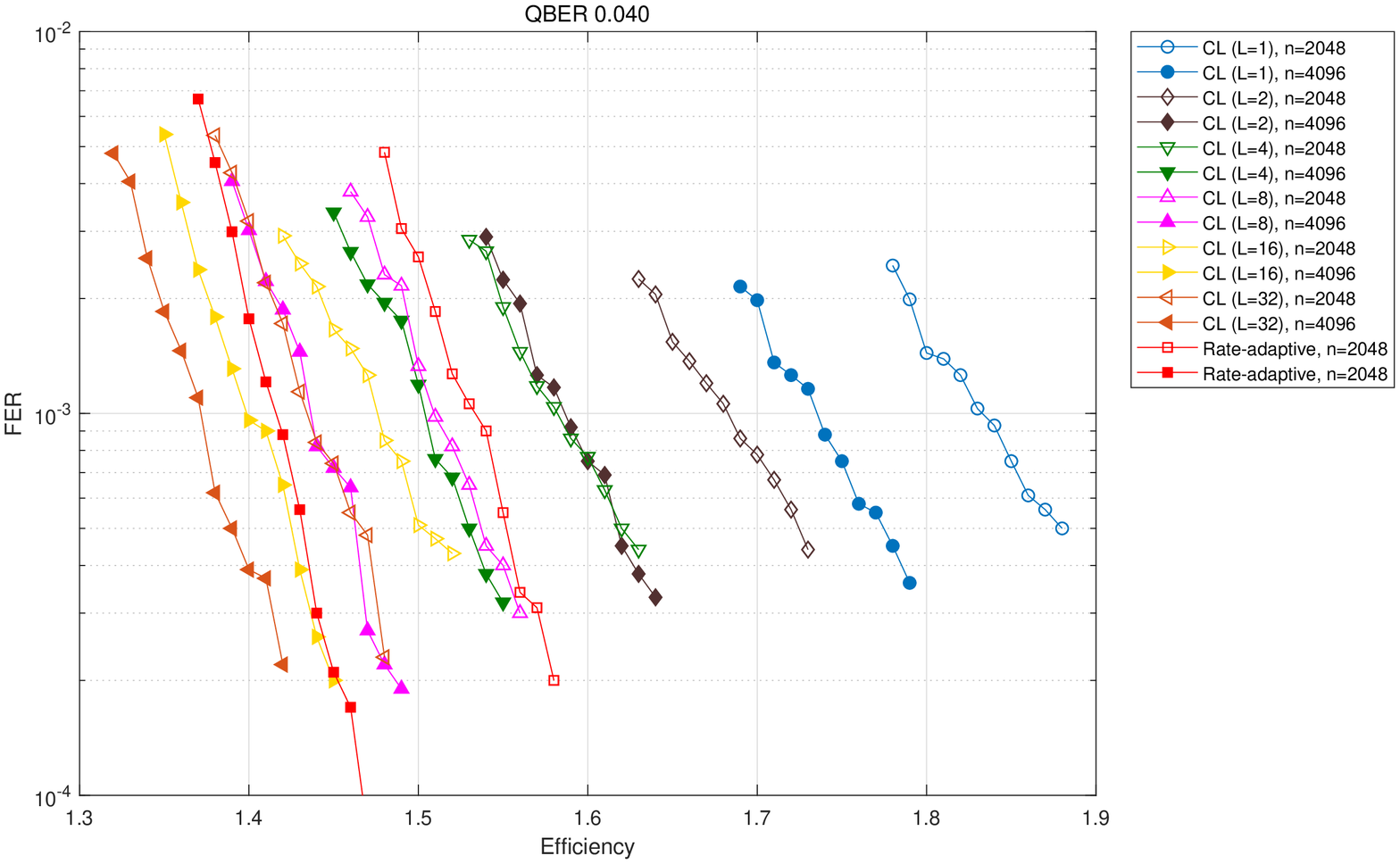, width=11cm}} 
\centerline{\epsfig{file=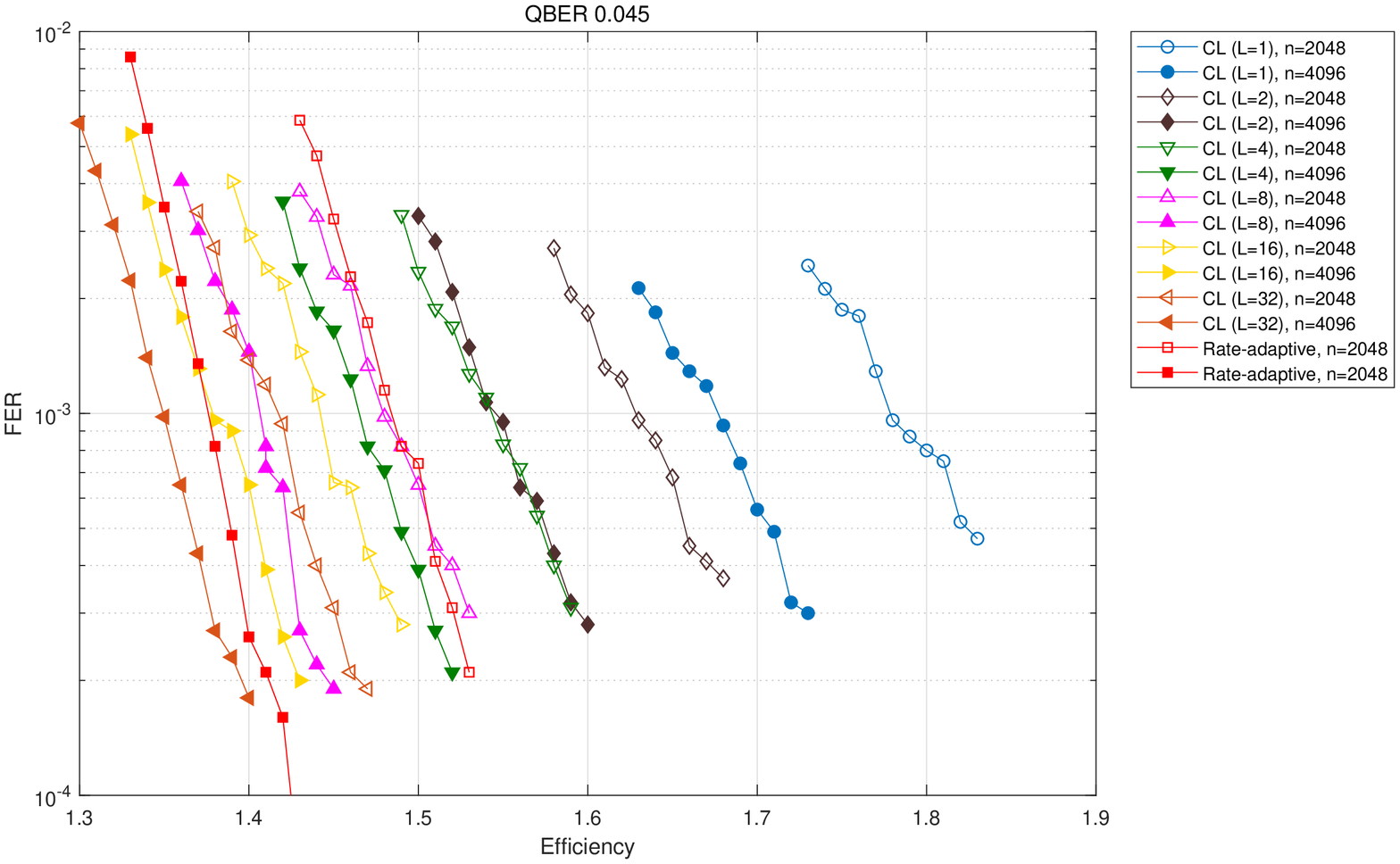, width=11cm}} 
\centerline{\epsfig{file=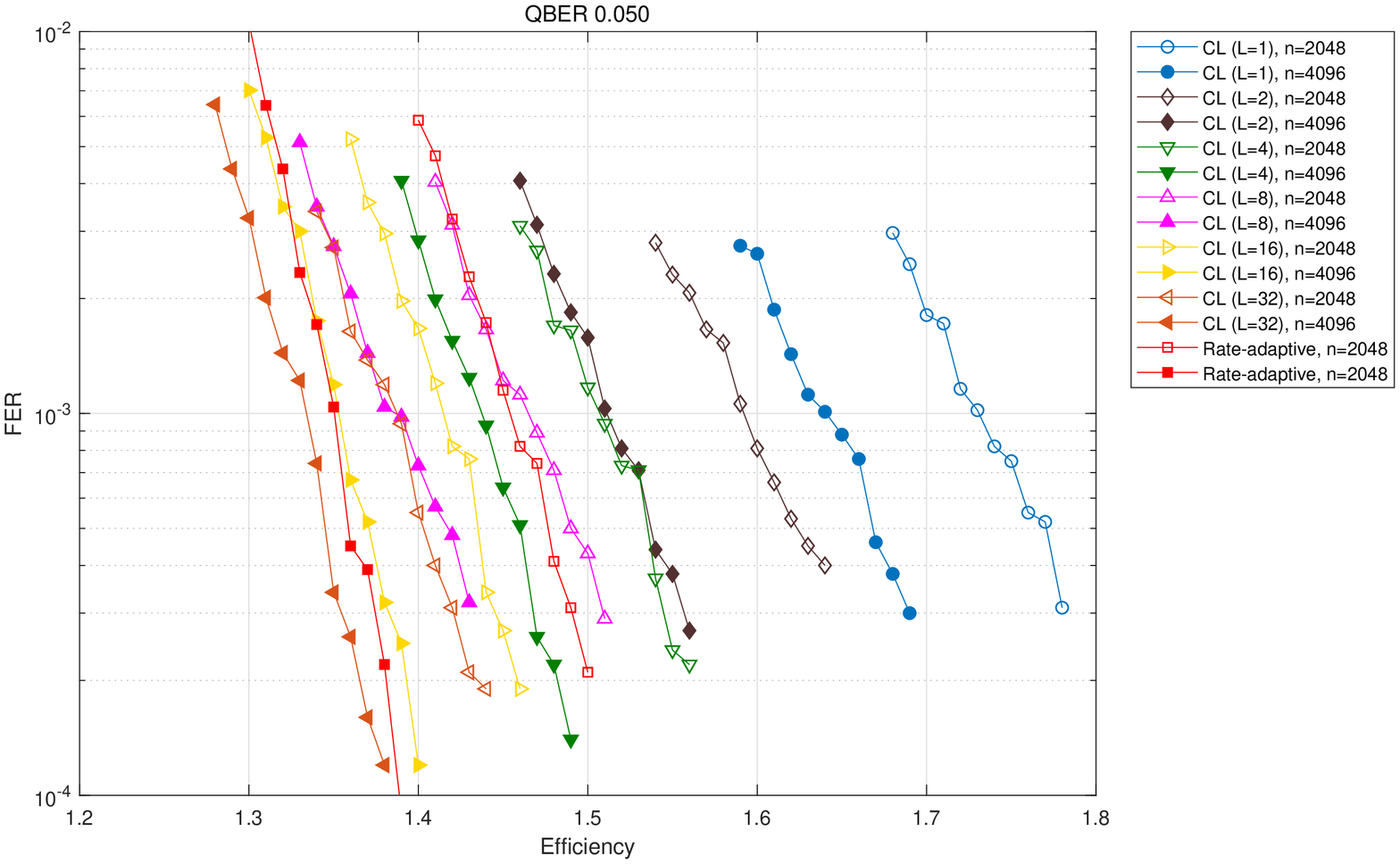, width=11cm}} 
\vspace*{13pt}
\fcaption{\label{fig:A3}FER curves in QBER of 0.040(top), 0.045(middle) and 0.050(bottom)}
\end{figure}

\end{document}